\documentclass[a4paper]{article}

\usepackage{INTERSPEECH2021}
\usepackage{graphicx,xcolor}
\usepackage{slashbox}
\usepackage{multirow}

\title{Large-Scale Pre-Training of End-to-End Multi-Talker ASR  for\\Meeting Transcription with Single Distant Microphone}
 
\name{Naoyuki Kanda, Guoli Ye, Yu Wu, Yashesh Gaur, Xiaofei Wang, Zhong Meng, \\Zhuo Chen, Takuya Yoshioka}
\address{
  Microsoft Corp., USA} 
\email{\{Naoyuki.Kanda,guoye,Wu.Yu,Yashesh.Gaur,Xiaofei.Wang,Zhong.Meng,zhuc,tayoshio\}@microsoft.com}

\begin{document}

\maketitle
\begin{abstract}
Transcribing meetings containing overlapped speech with only a single distant microphone (SDM) has been one of the most 
challenging problems for automatic speech recognition (ASR).
While various approaches have been proposed,
all previous studies on the monaural overlapped speech recognition problem 
were based on either simulation data or small-scale real data. 
In this paper, 
we extensively investigate a two-step approach 
where we first pre-train 
a serialized output training (SOT)-based multi-talker ASR 
by using large-scale simulation data 
and then fine-tune the model with a small amount of real meeting data.
Experiments are conducted by utilizing
75 thousand (K) hours of our internal single-talker
recording 
to simulate a total of 900K hours of 
multi-talker audio segments for supervised pre-training.
With fine-tuning on the 70 hours of the AMI-SDM training data,
our SOT ASR model 
achieves 
a word error rate (WER) of 21.2\% 
for the AMI-SDM evaluation set
while automatically counting speakers in each test segment.
This result 
 is not only significantly better than 
the previous state-of-the-art WER of 36.4\% with oracle utterance boundary information
but also better than
a result by a 
similarly fine-tuned single-talker ASR model applied to beamformed audio. 
\end{abstract}

\section{Introduction}

Meeting transcription with a distant microphone 
has been widely studied 
as 
one of the most 
challenging problems for automatic speech recognition (ASR) 
\cite{janin2003icsi,carletta2005ami,fiscus2007rich}.
The audio is noisy and reverberant due to the distance between the speaker and microphone and often includes overlapped utterances \cite{ccetin2006analysis}.
Meanwhile, the sentences are usually less grammatical in verbal communication, which creates additional difficulties for ASR.
While various approaches have been proposed 
especially for microphone array settings (e.g. \cite{yoshioka2019advances,watanabe2020chime}),
meeting transcription 
with only a single distant microphone (SDM)
is still highly challenging.
For example, the best reported word error rate (WER) 
for the AMI meeting corpus \cite{carletta2005ami}
is still over 35\% for the SDM setting even with oracle utterance boundary 
information
\cite{ko2017study,peddinti2017low,ganapathy20183,kanda2019acoustic}.

The meeting transcription system needs to 
recognize utterances from a variable number of speakers 
from audio which may contain overlapped utterances. 
To handle this, one approach is to 
first apply
speaker diarization  \cite{tranter2006overview,anguera2012speaker,park2021review}
to detect the utterance boundaries for each speaker and then perform ASR for each utterance. 
This method, however, could suffer from the accuracy degradation
in overlapped regions
because the ASR system is usually designed to recognize single-speaker speech. 
Another approach is to apply a speech separation system,
followed by the ASR system (e.g., \cite{chen2020continuous,raj2020integration}).
However, a speech separation system is usually trained 
with a signal-level criterion,
which is not necessarily optimal for ASR. 

To overcome this suboptimality, 
there has been a series of studies for multi-talker ASR that
 directly transcribes multiple utterances 
from overlapped speech. 
One popular approach is using a
neural network that has multiple output
layers, each of which recognizes one speaker 
\cite{yu2017recognizing,chang2019end,zhang2020improving,chang2020end,tripathi2020end,sklyar2020streaming,lu2020streaming}.
Permutation invariant training (PIT) \cite{yu2017permutation}
is usually used
to train such a multiple-output models.
One drawback of such
method is, however, that
the number of recognizable speakers 
is limited by the number of the output layers.
PIT also requires $O(S^3)$ of computation with respect to the number
of recognizable speakers $S$, 
which makes it inefficient to conduct large-scale training.
Recently,
serialized output training (SOT) was proposed
to recognize any number of utterances in speaker-overlapped audio \cite{kanda2020sot}.
The SOT-based ASR was shown 
to outperform PIT-based ASR
while automatically counting the speakers in the speaker-mixed audio. 
\cite{kanda2020sot} also showed that the SOT-based
ASR could be trained with $O(S)$ efficiency.

While promising results were shown, 
all the previous studies 
were limited to 
either simulated data \cite{yu2017recognizing,chang2019end,zhang2020improving,chang2020end,tripathi2020end,sklyar2020streaming,lu2020streaming,kanda2020sot,kanda2020joint,kanda2020investigation,chang2021hypothesis} or small-scale real data 
\cite{ko2017study,ganapathy20183,peddinti2017low,kanda2019acoustic}\footnote{Concurrently with our work, Chan et al. \cite{chan2021speechstew} proposed to mix various corpora to pre-train a single large ASR, showing a WER of 21.7\% for AMI-SDM evaluation set.}.
It is due to the difficulty in collecting real meeting recordings with precise transcriptions at large-scale.
One potential approach to the data scarcity problem
is large-scale pre-training, which has been studied
for {\it single-talker} ASR
(e.g., with labeled data 
   \cite{huang2014multi,joshi2020transfer}
  or with unlabeled data \cite{schneider2019wav2vec,baevski2020wav2vec}). 
However, it is still an open question if 
we can learn
a good representation for {\it multi-talker audio which is sometimes heavily overlapped}.

To further advance the SDM-based meeting transcription,
we extensively explore the 
supervised pre-training of SOT-based multi-talker ASR system with large-scale simulation.
Experiments are conducted with
75 thousand (K)
hours of single-talker data 
to simulate a total of 900K hours of multi-talker audio segments for pre-training.
For the AMI-SDM evaluation set, 
 the proposed multi-talker model with large-scale pre-training 
achieves 
a substantially better WER than the previously known results
after fine-tuning on the AMI-SDM training set. 

\section{SOT-Based Multi-Talker ASR}

\subsection{ASR based on  attention encoder decoder} 

Given input 
 $X\in\mathbb{R}^{f^a\times l^a}$, 
 where $f^a$ and $l^a$ are the feature dimension and the sequence length,
 respectively.
The goal of ASR system
  is 
  to estimate
 transcription $Y=(y_n\in \{1,...,|\mathcal{V}|\}|n=1,...,N)$,
where $|\mathcal{V}|$  is the size of the vocabulary $\mathcal{V}$, 
 and  $N$ is the number of estimated tokens.

In this paper, we use the attention-based encoder-decoder (AED)
\cite{chorowski2014end,chorowski2015attention}
as the backbone of the ASR system,
which is
represented as follows: 
 \begin{align}
 H &={\rm Encoder}(X),  \label{eq:enc}  \\
 o_n &= {\rm Decoder}(y_{[1:n-1]}, H).  \label{eq:asrout}
 \end{align}
The Encoder module
first converts $X$ 
into a sequence of hidden embeddings $H \in \mathbb{R}^{f^h\times l^h}$ for ASR (Eq. \eqref{eq:enc}),
where $f^h$ and $l^h$ are the embedding dimension and 
the sequence length, respectively.
At each decoder step $n$, 
the Decoder module 
calculates
the output distribution $o_n \in \mathbb{R}^{|\mathcal{V}|}$ 
 given
previous token estimates  $y_{[1:n-1]}$
and
$H$
 (Eq. \eqref{eq:asrout}).
The posterior probability
of token $i$ (i.e. the $i$-th token in the vocabulary $\mathcal{V}$) 
at the $n$-th decoder step 
is represented as
 \begin{align}
Pr(y_n=i|y_{[1:n-1]},X) = o_{n,i}, \label{eq:tokenprob}
\end{align}
where $o_{n,i}$ represents
the $i$-th element of $o_n$.
The posterior probability of token $Y$ given input $X$ 
is represented as, 
\begin{align}
Pr(Y|X) =&\prod_{n=1}^{N}Pr(y_{n}|y_{[1:n-1]}, X).
\end{align}

In this paper, we use a modified version of Conformer network \cite{gulati2020conformer}
for Encoder module, and 
a conventional Transformer-based decoder \cite{vaswani2017attention}
for Decoder module.
The modifications we have made to the Conformer network are as follows: 
 (i) we insert a squeeze and excitation module \cite{hu2018squeeze} 
  just before the dropout of the convolution module; 
  (ii) we do not use batch normalization in the convolution module; 
and (iii) we add one more point-wise convolution after 
depth-wise convolution.
These changes have been made based on our preliminary test.

\subsection{SOT for multi-talker ASR}
SOT was proposed for the AED to recognize a variable number of speakers from possibly overlapped audio \cite{kanda2020sot}.
In the SOT framework, 
the multiple utterances are concatenated to form a single token sequence
by inserting 
 a special symbol $\langle sc\rangle$
 representing a speaker change. 
For example, for the three-speaker case, a reference token sequence is given as
$R=\{r^1_{1},..,r^1_{N^1}, \langle sc\rangle, r^2_{1},..,r^2_{N^2}, \langle sc\rangle, r^3_{1},..,r^3_{N^3}, \langle eos\rangle\}$, 
where $r^j_i$ represents the $i$-th token of the $j$-th speaker.
Here, $\langle eos\rangle$, a token for sequence end,  is used only at the end of the entire sequence. 
In the inference, the decoding process is iterated until
$\langle eos\rangle$ is detected so that
 the SOT-based model can theoretically transcribe utterances of any number of speakers
while automatically estimating the number of speakers. 

There are multiple ways to determine the speaker order to form reference label sequence $R$.
One simple yet effective approach proposed in \cite{kanda2020sot} is
sorting the reference labels
by their start times, which is called ``first-in, first-out'' (FIFO) training.
FIFO training works with complexity of $O(S)$ 
with respect to the number of speakers $S$ while
showing superior accuracy than
 a scheme that exhaustively considers 
all possible permutations \cite{kanda2020sot}.
In this paper, we always use this FIFO training
scheme.

\begin{table}[t]
  \caption{Statistics of AMI corpus.}
  \label{tab:stats}
  \vspace{-3mm}
  \centering
  {\scriptsize
  \begin{tabular}{c|c|rrrr}
\multicolumn{6}{c}{(a) utterance} \\
    \toprule
& \multicolumn{1}{c|}{\# of} & \multicolumn{1}{c}{\# of} & \multicolumn{1}{c}{Average} & \multicolumn{1}{c}{Total}  & \multicolumn{1}{c}{\# of}     \\ 
&\multicolumn{1}{c|}{speakers} &  \multicolumn{1}{c}{segments} &  \multicolumn{1}{c}{dur. (sec)} & \multicolumn{1}{c}{dur. (hr)} & \multicolumn{1}{c}{words}   \\ \midrule
 train & 1 & 107,319  & 2.6 & 76.9  & 793,856 \\ 
dev & 1 & 13,098 & 2.5 & 8.9 & 94,953 \\ 
eval & 1 & 12,643 & 2.5 & 8.7 & 89,666 \\
    \bottomrule
\multicolumn{6}{c}{} \\
\multicolumn{6}{c}{(b) utterance group} \\
    \toprule
&  \multicolumn{1}{c|}{\# of} & \multicolumn{1}{c}{\# of} & \multicolumn{1}{c}{Average}  & \multicolumn{1}{c}{Total} &\multicolumn{1}{c}{\# of}     \\ 
& \multicolumn{1}{c|}{speakers} &  \multicolumn{1}{c}{segments} &  \multicolumn{1}{c}{dur. (sec)} & \multicolumn{1}{c}{dur. (hr)} & \multicolumn{1}{c}{words}   \\ \midrule
    & 1 & 33,646 & 3.0 & 28.5 & 298,157 \\
    &  2 & 12,914 & 5.5 & 19.6 & 239,820  \\
train & 3 & 5,450 & 8.2& 12.5 & 171,266 \\
    & 4 & 1,779 & 11.6 & 5.7 & 84,514 \\ 
    & 5 & 3 & 6.7 & 0.006 & 99 \\ \cmidrule{2-6}
& Total & 53,792 & 4.4 & 66.2& 793.856 \\
\midrule
    & 1 & 4,280 & 2.9 & 3.4 & 36,745 \\
    &  2 & 1,578 & 4.9 & 2.2 & 27,675 \\
dev & 3 & 680 & 7.3 & 1.4 & 19,895 \\
    & 4 & 254 & 9.5 & 0.7 & 10,638 \\ \cmidrule{2-6}
& Total & 6,792 & 4.1 & 7.6 & 94,953 \\
\midrule
     & 1 & 3,956 & 3.0 & 3.3 &  34,076 \\
     & 2 & 1,347 & 5.1 & 1.9 & 24,036 \\
eval & 3 & 631 & 8.0 & 1.4 &  20,276 \\
     & 4 & 203 & 13.2 & 0.7 & 11,278 \\ \cmidrule{2-6}
& Total & 6,137 & 4.3 & 7.3 & 89,666 \\
    \bottomrule
  \end{tabular}
  }
  \vspace{-5mm}
\end{table}

\begin{table*}[t]
  \caption{WER (\%) for AMI test set with various configurations. 
  Single Distant Microphone (SDM) evaluation is our primary focus in this paper, 
  and the results with Multiple Distant Microphone (MDM), i.e., 8-ch microphone array, are shown as reference numbers. A result with $^\ddagger$ is reported in a concurrent work in which a mixed corpus including AMI is used \cite{chan2021speechstew}.}
  \label{tab:overall}
  \vspace{-3mm}
  \centering
  {\scriptsize
  \begin{tabular}{c|c|ccc|c|c|cc}
    \toprule
Config. & Audio & \multicolumn{3}{c|}{ASR}  & Front-end & Evaluation & \multicolumn{2}{c}{WER (\%)} \\ 
ID & device & Architecture  & Pre-training & Fine-tuning &  & segment & dev & eval \\ \midrule \midrule
Kanda et al. \cite{kanda2019acoustic} & SDM &  CNN-TDNN-BLSTM hybrid      & - & AMI & -          & utterance   & 33.4 & 36.4 \\ 
Kanda et al. \cite{kanda2019acoustic} & MDM &  Multi-ch CNN-TDNN-BLSTM hybrid   & - & AMI & BeamFormIt & utterance   & 30.1 & 32.3 \\ 
Chan et al. \cite{chan2021speechstew} & SDM &  Conformer RNN-T      & Mixed (incl. AMI) & - & -          & utterance   & - & 21.7$^\ddagger$ \\ \midrule \midrule
1 & SDM &  Single-talker Conformer AED                & - & AMI & -            & utterance & 47.5  & 51.1 \\ 
2 & SDM &  Single-talker Conformer AED                & 75K & - & -            & utterance & 35.3 & 40.1 \\ 
3 & SDM &  Single-talker Conformer AED                & 75K  & AMI & -  & utterance & 23.0 & 25.8 \\  
4 & MDM &  Single-talker Conformer AED                & 75K  & AMI & BeamFormIt  & utterance & 22.2 & 24.4 \\  \midrule
5 & SDM &  SOT Multi-talker Conformer AED           & -   & AMI &-          & utterance group & 57.2  & 59.1 \\ 
6 & SDM &  SOT Multi-talker Conformer AED           & 75K &  -  &-          & utterance group & 41.6  & 44.2 \\ 
7 & SDM &  SOT Multi-talker Conformer AED           & 75K & AMI &-          & utterance group & {\bf 18.4} & {\bf 21.2} \\ 
    \bottomrule
  \end{tabular}
  } 
  
  \vspace{-5mm}
\end{table*}

\section{Experimental Settings}

\subsection{Data}
\subsubsection{Simulated multi-talker audio based on 75K-hour data}
\label{sec:simulation}

We used 
64 million 
 anonymized and transcribed English utterances,
totaling 75K hours,
for the data simulation to pre-train the multi-talker ASR models.
The data includes audio from various domains such as voice search and dictation.
Each audio is assumed to contain single speaker's voice.
Although the data could contain 
untranscribed interference speaker's speech in the background noise,
we did not apply any filtering.

For the pre-training, we simulated  multi-talker recordings on-the-fly
in our data loader module.
Specifically, we picked $N$ audio samples from the 75K-hour data by assuming they are from different speakers
and mixed them by adding delay to each sample, where $N$
was randomly chosen from 1 to 5.
The delay amount was also randomly sampled under the constraints that
there was at least 0.5 sec of difference in the start time of
each audio sample 
and that each audio sample had at least one overlapping region 
with another sample. 
After mixing the speech samples, we applied speed perturbation \cite{ko2015audio} of 
0.9--1.1x to further increase the data variation.
Thanks to the on-the-fly simulation,
 there was almost no duplication of simulated data through
 the entire pre-training process.

\subsubsection{AMI meeting corpus}

We used the AMI meeting corpus \cite{carletta2005ami}
for the evaluation as well as the fine-tuning of the pre-trained models.
The corpus comprises approximately 100 hours of meeting recordings, each containing 
three to five participants.
The audio
was recorded by an 8-ch microphone array, which is often called 
multiple distant microphone (MDM).
The first channel of the MDM audio is used for monaural ASR evaluation,
referred to as a single distant microphone (SDM) setting.
The AMI corpus
also contains the recordings from independent headset microphones (IHM)
worn by each participant.

As mentioned earlier, the primary focus of this paper is the evaluation with SDM recordings.
Meanwhile, we sometimes used the MDM or IHM recordings for analysis purposes.
 We used scripts in Kaldi toolkit \cite{povey2011kaldi} to
partition
the AMI corpus into training, development and evaluation recordings.
The statistics about the AMI meeting corpus 
 are shown in Table \ref{tab:stats}.
The definition of ``utterance group'' 
 in the table
 will be explained in the next section.

\begin{figure}[t]
  \centering
  \includegraphics[width=1.1\linewidth]{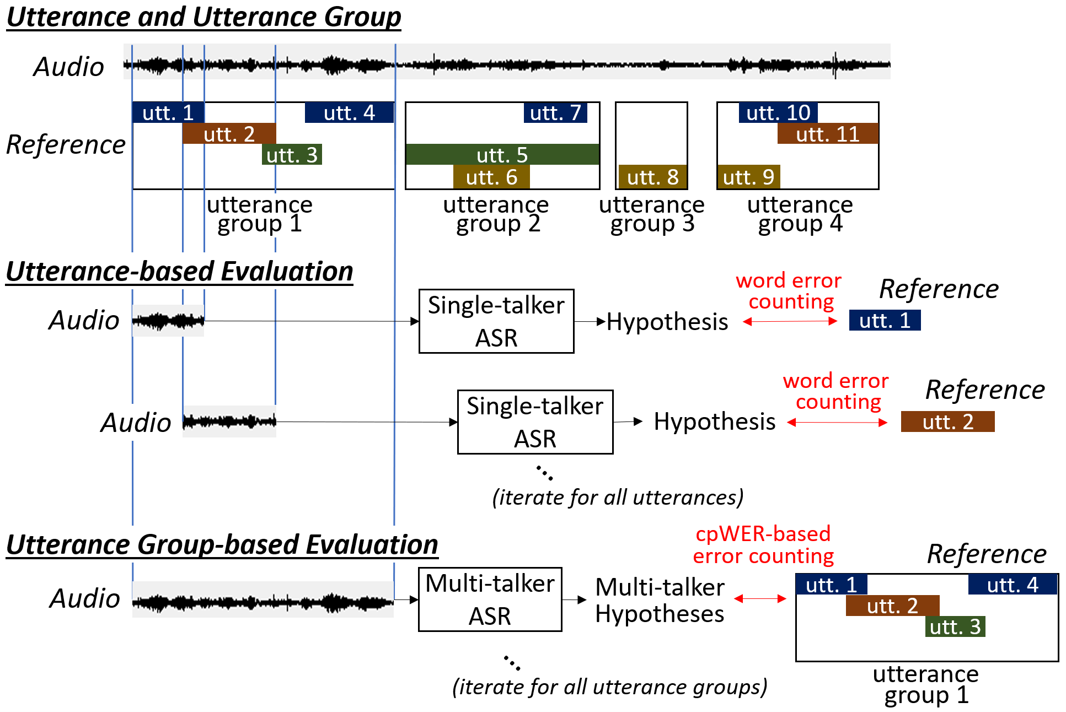}
  \vspace{-7mm}
      \caption{Utterance and utterance group for segmentation.}
  \label{fig:segments}
  \vspace{-5mm}
\end{figure}

\subsection{Evaluation metric}
\label{sec:metric}

In this paper, we introduce a notion of 
``utterance group''
to appropriately evaluate
 multi-talker ASR models.
The relationship between the utterance and utterance group is
illustrated in Fig. \ref{fig:segments} top. 
The utterance group is defined as a set of utterances
that are connected by speaker overlap regions.
In other words, the utterance groups can be formed
by segmenting the recording at silence positions or non-overlapping utterance boundaries.

{\it The utterance-based evaluation} is a standard way to
evaluate single-talker ASR models for AMI.
As shown in the middle of Fig. \ref{fig:segments},
ASR is applied to each speech segment obtained with reference  utterance boundaries.
 In the MDM setting, 
 beamforming is used 
 before the signals are fed into the single-talker ASR system.
Mis-recognized words are counted for each utterance,
 and the errors are summed up to calculate a final WER.

{\it The utterance group-based evaluation} 
is newly introduced in this paper 
to evaluate the multi-talker ASR models.
As shown in Fig. \ref{fig:segments} bottom,
multi-talker ASR is applied to each utterance group segment without information about the utterance boundaries inside the segment. 
The multi-talker ASR system generates 
hypotheses consisting of one or more utterances from the utterance group audio.
We calculate the number of mis-recognized words
based on the concatenated minimum-permutation word error rate \cite{watanabe2020chime}.
Specifically, we first concatenate the reference transcriptions
 of the same speaker in the utterance group (i.e., in the example of Fig. \ref{fig:segments}, utterance \#1 and \#4 
 in utterance group \#1).
Then,
 the best alignment between the hypotheses and 
 the concatenated references 
is selected 
among all possible permutations\footnote{If the number of hypotheses are larger than that of references,
all hypotheses that don't have corresponding references are counted
as insertion errors. Similarly, if the number of references are larger than 
that of hypotheses, all unmatched references are counted as deletion errors.} for error counting.
Finally, the errors from each utterance group are summed up
and divided by the number of total reference words
 to calculate the WER.
 Note that the denominator to calculate the WER, 
 i.e. the number of the total reference words, is the same for
the utterance-based evaluation
 and the utterance group-based evaluation.

\begin{table}[t]
  \caption{Speaker counting accuracy (\%) by SOT Multi-talker AED for each utterance group of AMI-SDM evaluation set.}
  \label{tab:spk-count}
  \vspace{-3mm}
  \centering
  {\scriptsize
  \begin{tabular}{c|c|cccccc}
    \toprule
Model & Actual \# of& \multicolumn{6}{c}{Estimated \# of Speakers (\%)} \\
      & Speakers & 0& 1 & 2 & 3 & 4 & $>$5 \\ \midrule
 Before      & 1 & 2.1 & {\bf 97.8} & 0.1 & 0.0 & 0.0 & 0.0 \\ 
 fine-tuning & 2 & 0.1 & 96.7 & {\bf 3.3} & 0.0 & 0.0 & 0.0 \\ 
 (Config. 6)      &     3 & 0.0 & 95.6 & 4.3 & {\bf 0.2} & 0.0 & 0.0 \\ 
             & 4 & 0.0 & 94.1 & 5.9 & 0.0 & {\bf 0.0} & 0.0 \\ \midrule
 After       & 1 & 0.2 & {\bf 97.2} & 2.5 & 0.1 & 0.0 & 0.0\\ 
 fine-tuning & 2 & 0.0 & 13.7 & {\bf 80.5} & 5.9 & 0.0 & 0.0\\ 
 (Config. 7) & 3 & 0.0 & 2.4 & 32.6 & {\bf 60.2} & 4.8 & 0.0\\ 
             & 4 & 0.0 & 0.0 & 9.9 & 51.2  & {\bf 38.9} & 0.0 \\ 
    \bottomrule
  \end{tabular}
  }
  \vspace{-3mm}
\end{table}

\begin{table}[t]
  \caption{WER (\%) for AMI-SDM evaluation set by SOT Multi-talker AED w.r.t the number of speakers in utterance group.}
  \label{tab:detail}
  \vspace{-3mm}
  \centering
  {\scriptsize
  \begin{tabular}{c|cc|cccc|c}
    \toprule
& Pre- & Fine-  & \multicolumn{4}{c|}{\# of speakers in the segment} & Total\\ 
& training & tuning &     1& 2 & 3 & 4 & \\ \midrule
Config. 5 &- & AMI  & 37.8  & 59.5 & 76.8  & 88.8 & 59.1\\
Config. 6 & 75K & -  & 22.8  & 42.0  & 62.7  & 77.4  & 44.2 \\
Config. 7 & 75K & AMI  &   14.7 & 19.6 & 25.7 & 35.5 & 21.2 \\
    \bottomrule
  \end{tabular}
  }
  \vspace{-6mm}
\end{table}

\subsection{ASR model configuration}

In our investigation, we 
trained single-talker ASR models as well as 
SOT-based multi-talker ASR models.
We used the same encoder-decoder architecture for all cases.
Specifically, the encoder consisted of 
2 layers of convolution layers 
that subsample the time frames by a factor of 4,
followed by
18 conformer layers.
Each conformer layer consisted of
two 1024-dim feed forward layers in a sandwitch structure,
a multi-head attention with 8 heads,
a depth-wise convolution with kernel size 3,
and a squeeze-and-excitation network with reduction factor 8
\cite{hu2018squeeze}.
The embedding dimension was set to 512.
The decoder consisted of
6 layers, each of which had a
multi-head attention with 8 heads and a 2048-dim
feed forward layer.
4K subwords \cite{kudo2018subword}
were used as a recognition unit.
We used a 80-dim log mel filterbank extracted 
every 10 msec for the input feature.

The single-talker model was pre-trained by the 75K-hour data
while the multi-talker model was pre-trained by the 75K-hour-based simulated multi-talker data. 
For both models,
we performed 425k training iterations with
32 GPUs, each of which consumed mini-batches of 24,000 frames (roughly corresponding to 
900K hours of data simulation and consumption).
We used Adam optimizer with
a linear decay learning rate schedule with a peak learning rate of 1e-3 
after 25k warm up iterations.
In the fine-tuning stage,
the pre-trained single-talker model was fine-tuned by the AMI-SDM {\it utterance} segments while
the pre-trained multi-talker model was fine-tuned by the AMI-SDM {\it utterance group} segments
with the speaker-based FIFO training scheme \cite{kanda2020investigation}.
For both cases, we used speed perturbation 
and SpecAugment \cite{park2019specaugment}, and
conducted 25k training iterations with 16 GPUs, each of which
consumed mini-batches of 6,000 frames. 
A linear decay learning rate schedule starting at a learning rate of 1e-4 was used. 
Besides the pre-training-based two-stage approach, 
we also trained models on the AMI-SDM data without pre-training.
In this case,
we used mini-batches of 6,000 frames and 
trained the models for 110k iterations with 16 GPUs, with 
a linear decay learning rate schedule with a peak learning rate of 1e-4 
after 10k warm up iterations.

\section{Evaluation Results}

\subsection{Main results}

Table \ref{tab:overall} shows our main results comparing
single-talker ASR and SOT-based multi-talker ASR models.
The two-stage approach of performing both pre-training and fine-tuning 
significantly improved the WER for both single-talker
(Config. 3) and multi-talker models (Config. 7).
The two-stage models outperformed 
both the
75K-hour-only models and AMI-only models with significant margins.
Interestingly, 
although the single-talker model outperformed the multi-talker model
before fine-tuning (Configs. 2 and 6),
the multi-talker model showed superior performance 
after fine-tuning (Configs. 3 and 7).
Table \ref{tab:spk-count} shows that 
fine-tuning significantly improved the 
speaker counting accuracy
of the multi-talker model. 
This improvement in
speaker counting led to the larger WER improvement of  
the multi-talker model
especially for 
segments with many speakers as shown in Table \ref{tab:detail} 
(Config. 6 vs. Config. 7).
Table \ref{tab:detail} also shows the
improvement provided by the pre-training
(Config. 5 vs. Config. 7),
where
large gains were observed 
for segments with more speakers.
This is because such segments were more scarce in the real data as 
shown in Table \ref{tab:stats} (b).

While our main focus in this paper is SDM-based recognition,
we also evaluated the fine-tuned single-talker ASR model for the MDM setting using BeamFormIt beamformer \cite{anguera2007acoustic}.
The result is shown in Config. 4 of Table \ref{tab:overall}.
We observed that the proposed SOT multi-talker AED model (Config. 7) 
outperformed the combination of the single-talker AED and MDM-based beamforming
while the SOT multi-talker AED model even does not require
the utterance-level boundary information.
This result shows the effectiveness of the end-to-end multi-talker ASR model
that jointly performs speaker counting, speech separation, and speech recognition.

\subsection{Why fine-tuning by real data is effective}

To better understand why fine-tuning worked so effectively,
we conducted an experiment using the AMI-IHM training data for fine-tuning.
Table \ref{tab:fine-tuning} shows the results. 
Here, we prepared two new datasets for fine-tuning: ``IHM real-mix'' and
``IHM rand-mix''.
``IHM real-mix'' was generated by first mixing all IHM recordings in the same meeting 
and then segmenting the mixed audio by the utterance groups. 
The difference from ``SDM'' to ``IHM real-mix'' 
lies only in 
the noise and reverberation characteristics.
On the other hand, 
``IHM rand-mix'' was generated by randomly mixing the AMI IHM utterances for up to 4 utterances.
The difference between ``IHM real-mix'' and ``IHM rand-mix''
resides in the speaker overlap pattern.
The biggest WER difference was observed
 between the no-fine-tuning (1st row) and the fine-tuning 
by ``IHM rand-mix'' (2nd raw),
which suggested
the fine-tuning of
the language model (LM), English accent,
and the accurate overlap of speech region\footnote{
In pre-training,
the simulated mixed audio may not have
the overlap of speech region due to the silence of the original audio. 
Also, 
the audio used in the simulation could contain interference speech 
without transcription.
We think these factors may be also refined by fine-tuning.} 
had the biggest impact.
The second largest WER difference was observed between ``IHM rand-mix''-based fine-tuning and 
``IHM real-mix''-based fine-tuning, 
showing the importance of including the real overlapping pattern. 
The difference from ``IHM real-mix'' to ``SDM''
was relatively small 
presumably because the pre-trained representation was already robust to noisy samples.

\subsection{Gap between SDM and IHM-based recognition}

We further conducted the evaluation to understand the 
gap 
between SDM-based recognition and IHM-based recognition,
the results of which are shown in
Table \ref{tab:oracle-eval}.
We prepared two additional evaluation sets, ``IHM'' and ``IHM-mix''.
 ``IHM-mix'' was generated by first mixing all IHM recordings in the same session 
 and then segmenting the mixed audio based on the utterance group.
The difference between ``SDM'' and ``IHM-mix'' is whether the audio is recorded 
by SDM or IHM.
On the other hand, the difference between ``IHM-mix'' and ``IHM'' 
includes whether the speech is overlapped and
whether we use additional utterance boundary information.
While we observed a significant gap between 
SDM and IHM-mix, the WER difference
between IHM-mix and IHM was relatively small. 
This indicates that the SOT AED
already worked 
well in terms of
speaker counting and multi-talker transcription
and that the refinement of the model capacity or the data simulation
is rather necessary
to better cope with noisier data.

\begin{table}[t]
\setlength{\tabcolsep}{1mm}
  \caption{WER (\%) of AMI-SDM test set by SOT Multi-talker AED with different fine-tuning data.}
  \label{tab:fine-tuning}
  \vspace{-3mm}
  \centering
  {\scriptsize
  \begin{tabular}{c|ccc|cc}
    \toprule
Training data   &LM, accent, \& & Overlap pattern& Noise \& reverb. & & \\ 
for fine-tuning & accurate overlap &  &  & dev & eval   \\ \midrule
-             &    -     &   -     &   -  & 41.6    & 44.2\\
IHM rand-mix & $\surd$ &   -     &   -   & 25.0   & 29.0 \\
IHM real-mix   & $\surd$ & $\surd$ &   -  & 19.6  & 23.0 \\
SDM            & $\surd$ & $\surd$ & $\surd$& 18.4 & 21.2 \\
    \bottomrule
  \end{tabular}
  }
  \vspace{-3mm}
\end{table}

\begin{table}[t]
\setlength{\tabcolsep}{1mm}
  \caption{WERs (\%) from IHM to SDM settings. 
 The single-talker AED was used for the utterance evaluation
 while the multi-talker AED was used for the utterance group evaluation.} 
  \label{tab:oracle-eval}
  \vspace{-3mm}
  \centering
  {\scriptsize
  \begin{tabular}{cc|ccc|cc}
    \toprule
Evaluation & Evaluation & Microphone & Speech & Utterance  & \multicolumn{2}{c}{WER (\%)} \\ 
  data    & segment &  distance?   & overlapped?     &  boundary?  & dev & eval \\ \midrule
 IHM & utterance & close & no & given   & 12.8 &  12.2  \\
 IHM-mix & utterance group & close & yes & n/a   & 13.5 & 14.9  \\ 
 SDM & utterance group & distant & yes & n/a   & 18.4 & 21.2\\
    \bottomrule
  \end{tabular}
  }
  \vspace{-5mm}
\end{table}

\section{Conclusions}

In this paper, 
we extensively explored the 
pre-training of SOT-based multi-talker ASR with large-scale simulation.
In the evaluation,  
the proposed multi-talker model 
achieved state-of-the-art WER of 21.2\% for AMI-SDM evaluation set,
even outperforming
 the combination of
MDM-based speech enhancement and strong single-talker ASR.

\bibliographystyle{IEEEtran}

\bibliography{mybib}

\end{document}